\documentstyle[12pt]{article}

\newcommand{\newc}{\newcommand}
\newc{\vrel}{v_{\rm rel}}
\newc{\vcm}{v_{\rm cm}}
\newc{\acm} {a_{\rm cm}}
\newc{\bcm} {b_{\rm cm}}
\newc{\ccm} {c_{\rm cm}}
\newc{\arel} {a_{\rm rel}}
\newc{\brel} {b_{\rm rel}}
\newc{\crel} {c_{\rm rel}}
\def\NPB#1#2#3{Nucl. Phys. {\bf B#1} (19#2) #3}
\def\PLB#1#2#3{Phys. Lett. {\bf B#1} (19#2) #3}

\def\PRD#1#2#3{Phys. Rev. {\bf D#1} (19#2) #3}
\def\PRL#1#2#3{Phys. Rev. Lett. {\bf#1} (19#2) #3}

\def\beq{\begin{equation}}
\def\eeq{\end{equation}}
\def\bea{\begin{eqnarray}}
\def\eea{\end{eqnarray}}

\begin{document}
\setlength{\baselineskip}{0.2in}

\rightline{UM-TH-94-10}
\rightline{March 1994}
\centerline{}
\vskip 1.5truein
\centerline{\large\bf Annihilation Cross Sections for Relic Densities}
\centerline{\large\bf in the Low Velocity Limit}
\bigskip
\bigskip
\bigskip
\centerline{James D. Wells\footnote{Email: {\tt jwells@umich.edu}}}
\centerline{Department of Physics}
\centerline{University of Michigan}
\centerline{Ann Arbor, Michigan 48109--1120}
\bigskip
\bigskip
\bigskip
\bigskip
\bigskip
\centerline{\bf Abstract}
A simple method to calculate the total annihilation cross section
of massive cold relics is presented.  Exact expressions for $\arel$,
$\brel$, and $\crel$ of $\sigma \vrel=\arel+\brel \vrel^2+\crel \vrel^4$
are obtained from simple derivatives of the squared annihilation amplitude.
Three examples are given.  In one example, the computed annihilation cross
section of massive Dirac neutrinos confirms recent correct calculations.

\vfill\eject
\setlength{\baselineskip}{0.3in}

\section{Introduction}

The number density of cold relics from the early universe depends
sensitively on the annihilation cross section of such particles.  The
equation governing the evolution of the number density
is the Boltzmann rate equation~\cite{lee},
\beq
\frac{dn}{dt}= -3 H n-<\sigma \vrel > (n^2-n_0^2)
\eeq
where $n$ is the number density, $H$ is the Hubble constant, $n_0$
is the number density in thermal equilibrium and $< \sigma\vrel >$
is the thermal averaged annihilation cross section.  The
thermal averaged cross section~\footnote{For a precise treatment of
the thermal average see refs.~\cite{srednicki} and~\cite{gondola}}
is approximately~\cite{an}
\beq
\label{tavg}
<\sigma \vrel >=\frac{\int d\vrel \vrel^2 (\sigma\vrel ) e^{-\vrel^2/4x}}
                   {d\vrel \vrel^2 e^{-\vrel^2/4x}}
\eeq
where $x=T/m$ (the cold relic mass is denoted as $m$).
The freeze-out temperature is defined as the temperature at which
the expansion
rate of a co-moving volume becomes larger than the rate of annihilation.
After freeze-out
the relic particle decouples from the background, and the subsequent
relic density is determined solely by the expansion rate of the
universe.

As has been shown by many authors the freeze-out temperature can
be very small so as to justify using the Boltzmann distribution
to calculate the thermal average.
Most notably this is true in the case of stable
neutralinos where $x_f\approx 1/20$~\cite{srednicki}.
This low decoupling energy leads one to consider
a low velocity expansion of the annihilation
cross section:
\beq
\label{sv1}
\sigma\vrel=\arel+\brel \vrel^2+\crel \vrel^4+\cdots .
\eeq
Obtaining the thermal averaged annihilation cross section then becomes
merely a calculation of \hbox{$<\vrel^{2n}>$}.
However, when the relic particles
annihilate into a narrow resonance special care must be
taken~\cite{kane}.  Although the low velocity expansion
breaks down under these conditions~\cite{gondola,griest1},
in most cases approximating the annihilation
cross section with Eq.~(\ref{sv1}) is extremely useful.
A crucial task remaining is the determination of
$\arel$, $\brel$, and $\crel$.

In this letter I will demonstrate a simple technique to calculate
the low velocity expansion of the annihilation cross section
of $2\to 2$ processes with arbitrary final state masses.  The technique
{\em substantially} increases the facility of annihilation cross section
calculations.
Instead of performing integrals to obtain $\sigma \vrel$ and then
expanding in $\vrel$, I expand in $\vrel$ and then do the remaining
integrals over simple trigonometric functions analytically
to arrive at the final answer.
This technique is made possible by the fact that the Mandelstam
variables have no angular dependence in the zero velocity limit.
Therefore, the differential cross section can have no angular
dependence in the zero velocity limit, and can be expanded
in a Taylor series with each power of velocity having coefficients
of simple trigonometric functions.  These simple trigonometric
functions are then integrated out painlessly to obtain the
total annihilation cross section.

\section{Expansion of the Annihilation Cross Section}

In this section the low velocity expansion of the $2\to 2$ annihilation
cross section with arbitrary final state masses is presented.
The incoming momenta are labelled
$p_1$ and $p_2$, both with mass $m$; the outgoing momenta
are labelled $p_3$ and $p_4$ with masses $m_3$ and $m_4$
respectively; and, $\vcm$ is the center of mass velocity of the
annihilating particles ($\vcm=\vrel/2$).
The differential cross section is
\beq
\frac{d\sigma}{d\Omega}=\frac{|M|^2}{64\pi^2 s}
\frac{|{\bf p_3}|}{|{\bf p_1}|}.
\eeq
First, the following definitions are useful
\footnote{Recall that $u=\sum m_i^2-s-t$ .}
\beq
J(s,t)={|M|^2},~~~{\rm and}~~~K(s)=\frac{|{\bf p_3}|}{16 \pi m s}.
\eeq
Since
\beq
|{\bf p_1}|=\frac{m \vcm}{\sqrt{1-\vcm^2}},~~~{\rm and}~~~
|{\bf p_3}|=\frac{\sqrt{ (s-(m_3+m_4)^2) (s-(m_3-m_4)^2) }}{2\sqrt{s}}
\eeq
the differential cross section can be written as
\beq
\label{diffcs}
\vcm \frac{d\sigma}{d\Omega}=\frac{J(s,t)}{4\pi} K(s) \sqrt{1-\vcm^2}.
\eeq

The low velocity expansion of the Mandelstam variables is
straightforward:
\bea
s & = & s_0+s_2 \vcm^2+s_4 \vcm^4+{\cal O} (\vcm^6) \\
t & = & t_0+t_1 \cos\theta \vcm+t_2 \vcm^2
   +t_3\cos\theta \vcm^3+t_4 \vcm^4 +{\cal O} (\vcm^5)
\eea
where
\bea
& & s_0 = s_2=s_4=4m^2 \\
& & t_0= \frac{-2m^2+m_3^2+m_4^2}{2}  \\
& & t_1=\frac{\mu_1^2}{2}  \\
& & t_2= -2m^2  \\
& & t_3=\frac{2m^2(4m^2-m_3^2-m_4^2)}{\mu_1^2}  \\
& & t_4= -2m^2
\eea
with
\beq
\mu_1^2=\sqrt{(4m^2-(m_3-m_4)^2)(4m^2-(m_3+m_4)^2)}.
\eeq
Next, expand $K(s)$ in a Taylor series:
\beq
K(s)=K_0+K_2 \vcm^2+K_4 \vcm^4+{\cal O} (\vcm^6)
\eeq
where
\bea
\label{k0eq}
& & K_0=\frac{\mu_1^2}{256\pi m^4}  \\
\label{k2eq}
& & K_2=\frac{\mu_1^2}{256\pi m^4}
  \left( -\frac{3}{2}+\frac{2m^2\mu_2^2}{\mu_1^4} \right)  \\
\label{k4eq}
& & K_4=\frac{\mu_1^2}{256\pi m^4}
   \left( \frac{3}{8}+\frac{8m^4}{\mu_1^4}-\frac{m^2\mu_2^2}{\mu_1^4}
   -\frac{2m^4\mu_2^4}{\mu_1^8} \right)
\eea
with $\mu_2^2$ defined to be
\beq
\mu_2^2=8m^2-(m_3-m_4)^2-(m_3+m_4)^2.
\eeq

Expanding $J(s,t)/ 4\pi$ about $s=s_0$ and $t=t_0$ yields
\bea
\frac{J(s,t)}{4\pi} &= & \frac{1}{4\pi}
  \left( J_0+\frac{\partial J_0}{\partial s}(s-s_0)
  + \frac{\partial J_0}{\partial t} (t-t_0) \right. \nonumber \\
& & \left. + \frac{1}{2} \frac{\partial^2 J_0}{\partial s^2} (s-s_0)^2
  +\frac{1}{2} \frac{\partial^2 J_0}{\partial t^2} (t-t_0)^2
  + \frac{\partial^2 J_0}{\partial s\partial t}
         (s-s_0)(t-t_0) +\cdots \right)
\eea
where $J_0\equiv J(s_0,t_0)$.
The key point here is that $J_0$ and
$\partial^{m+n} J_0/\partial s^m \partial t^n$ do not depend on
$\cos\theta$; only $(t-t_0)^n$ can
depend on $\cos\theta$.  Therefore, to fourth order in velocity only the
functions $\cos^n\theta$ with $n=0\ldots 4$ need to be integrated.
Of course $\int d\Omega \cos\theta=\int d\Omega\cos^3\theta=0$
which forces all terms in $\sigma \vrel$ which are odd powers of $\vrel$
to be zero.  The result of integrating $J(s,t)/4\pi$ in the
low velocity limit is
\beq
\int d\Omega \frac{J(s,t)}{4\pi}=J_0+J_2 \vcm^2+J_4 \vcm^4+{\cal O} (\vcm^6)
\eeq
where
\bea
\label{j0eq}
J_0 &= &J(s_0,t_0) , \\
\label{j2eq}
J_2 & = &s_2\frac{\partial J_0}{\partial s}
    +t_2\frac{\partial J_0}{\partial t}
    +\frac{1}{6} t_1^2\frac{\partial^2 J_0}{\partial t^2} ,  \\
\label{j4eq}
J_4 &= & t_4\frac{\partial J_0}{\partial t}
   +\frac{1}{2}t_2^2\frac{\partial^2 J_0}{\partial t^2}
   +\frac{1}{3} t_1 t_3\frac{\partial^2 J_0}{\partial t^2} \nonumber \\
& &  + \frac{1}{6} t_1^2 t_2\frac{\partial^3 J_0}{\partial t^3}
  +\frac{1}{120} t_1^4\frac{\partial^4 J_0}{\partial t^4}
  +s_4\frac{\partial J_0}{\partial s}  \nonumber \\
& &  +s_2 t_2 \frac{\partial^2 J_0}{\partial s\partial t}
  +\frac{1}{6} s_2 t_1^2\frac{\partial^3 J_0}{\partial s\partial t^2}
  +\frac{1}{2} s_2^2\frac{\partial^2 J_0}{\partial s^2}.
\eea

Expanding out all the terms in Eq.~(\ref{diffcs}) the total annihilation
cross section to fourth order in velocity is
\beq
\label{eq1}
\sigma \vcm=\acm+\bcm \vcm^2+\ccm \vcm^4
\eeq
where
\bea
\label{areleq}
& & \acm = \frac{1}{2}\arel=J_0 K_0 \\
\label{breleq}
& & \bcm = 2\brel=J_0 K_2-\frac{1}{2} J_0 K_0+J_2 K_0  \\
& & \ccm = 8\crel=
          J_0 K_4-\frac{1}{2} J_0 K_2-\frac{1}{8} J_0 K_0+J_2 K_2
  -\frac{1}{2} J_2 K_0+J_4 K_0.
\eea

Thus, in principle the
coefficients $\acm$, $\bcm$, and $\ccm$ are easily calculated once the
squared amplitude for the annihilation process is known.  Although
the derivatives of $J(s,t)$ can be quite complicated, they are
nevertheless always tractable.  One is now faced only with the algebraic
problem of taking derivatives.

\section{Applications of the General Formula}
\bigskip

\subsection{Annihilation of Massive Neutrinos}
\bigskip

Limits on the mass of neutrinos can be inferred from requiring
that their relic density not overclose the universe.  These
bounds depend sensitively on their annihilation cross section.
As a check to the above formalism, I calculate the annihilation
cross section of both Majorana neutrinos and Dirac neutrinos
into fermions and compare with previous calculations in the literature.
In the case of Dirac neutrinos I confirm a recent result~\cite{dolgov}
which disagrees with earlier literature.

The interaction Lagrangian of Majorana neutrino annihilations
into fermions is
\beq
{\cal L}=\frac{G_F}{\sqrt{2}} {\bar \nu}\gamma^\mu\gamma_5 \nu
  {\bar f} \gamma_\mu (C_V-C_A\gamma_5) f
\eeq
which corresponds to the following squared amplitude:
\bea
J(s,t)& =& 4 G_F^2 (C_A^2+C_V^2) [ 2m^4+2m_f^4
    -4m^2s+s^2-4m^2 t-4m_f^2t+2st+2t^2] \\
& & +16 G_F^2 \left[
   (5 C_A^2-C_V^2) m^2 m_f^2-C_A^2 m_f^2 s\right]  \nonumber
\eea
where $m$ is the mass of the neutrino and $m_f$ is the mass of the
fermion.

Using the technique developed in the previous section, the
values of $\arel$ and $\brel$ are easily calculated to give
the following annihilation cross section to ${\cal O} (\vrel^2)$:
\bea
\label{maj}
\sigma\vrel &= & \arel + \brel\vrel^2  \nonumber \\
 &= & \frac{2G_F^2}{\pi} m^2 \sqrt{1-m_f^2/m^2}
   \left\{ (C_V^2+C_A^2) \frac{\vrel^2}{6} \right. \nonumber \\
  & & \left. + \frac{C_A^2}{2}\frac{m_f^2}{m^2}
     +\left[ C_V^2-\frac{17}{4} C_A^2+
     \frac{3}{4} \frac{C_A^2}{(1-m_f^2/m^2)} \right]
   \frac{m_f^2}{m^2}\frac{\vrel^2}{12} \right\}
\eea
The above equation is identical to the annihilation cross section calculated
by ref.~\cite{kolb} and others.

The Dirac neutrino annihilation cross section is calculated
similarly.  The interaction Lagrangian is
\beq
{\cal L}=\frac{G_F}{\sqrt{2}} {\bar\nu}\gamma^{\mu} (1-\gamma_5 )\nu
{\bar f}\gamma_\mu (C_V-C_A\gamma_5 ) f
\eeq
which gives the following squared amplitude:
\bea
J(s,t) &= & 4 G_F^2 (C_A^2+C_V^2) (m^4+m_f^4-2m^2 t-2m_f^2 t+t^2) \nonumber \\
& & +2 G_F^2 (C_V-C_A)^2 (s^2+2st-2m^2s) \nonumber \\
& & +8 G_F^2 (2 C_A^2 m^2 m_f^2-C_A^2 m_f^2 s+C_A C_V m_f^2 s).
\eea
The low velocity expansion of $\sigma\vrel$ then becomes
\bea
\sigma\vrel & = & \frac{G_F^2}{2\pi} m^2 \sqrt{1-m_f^2/m^2}
     \left\{ (C_V^2+C_A^2) \left[ 1+\frac{\vrel^2}{24}
      \left( 5-2\frac{m_f^2}{m^2}+\frac{3}{1-m_f^2/m^2}
       \right) \right] \right. \nonumber \\
  & & \left. +\frac{1}{2} (C_V^2-C_A^2) \frac{m_f^2}{m^2}
     \left[ 1+\frac{\vrel^2}{8} \left( 1 +\frac{1}{ 1-m_f^2/m^2}
      \right) \right] \right\} .
\eea
This result differs from refs.~\cite{kolb,kane} but confirms
the recently published calculation of Dolgov
and Rothstein~\cite{dolgov}.

\subsection{Neutralino s--channel Annihilation}
\bigskip

Ellis {\it et al.}~\cite{ellis} calculated the thermal averaged
cross section to ${\cal O}(x)$
for the s--channel Z boson
contribution to neutralinos annihilating into fermions.
As a final example to demonstrate the technique developed
in this paper, I calculate the annihilation cross section
and compare.
Following the
conventions of ref.~\cite{ellis} the
interaction Lagrangian is
\beq
{\cal L}=\sum_f ({\bar \chi}\gamma^\mu\gamma_5\chi )
 \frac{ (g_{\mu\nu}-q_\mu q_\nu /m_Z^2) m_Z^2}{s-m_Z^2+i\Gamma_Z m_Z}
	[ {\bar f}\gamma^\nu (A_Z P_L+B_Z P_R) f].
\eeq

{}From this Lagrangian the squared amplitude is calculated, and
the values of $J_0$, $J_2$, $K_0$, and $K_2$ are determined.
Using Eq.~(\ref{tavg}) I find the coefficients $a_x$ and $b_x$
of $<\sigma\vrel >=a_x+b_x x$ to be
\bea
\label{atilanswer}
a_x & = &\sum_f \frac{c_f}{2\pi} \sqrt{1-\frac{m_f^2}{m^2}}
   \frac{m_Z^4}{(m_Z^2-4 m^2)^2+\Gamma_Z^2 m_Z^2} m_f^2 (A_Z-B_Z)^2
  \left[ 1-4(\frac{m}{m_Z})^2 \right]^2 , \\
b_x &=& \sum_f \frac{c_f}{2\pi} \sqrt{1-\frac{m_f^2}{m^2}}
   \left. \frac{m_Z^4}{(m_Z^2-4 m^2)^2+\Gamma_Z^2 m_Z^2}
  \right\{ 4m^2(A_Z^2+B_Z^2)  \nonumber \\
 & & -4m_f^2
  \left[ A_Z^2-3A_ZB_Z+B_Z^2-3\left( \frac{m}{m_Z} \right)^2
  (A_Z-B_Z)^2 \right] \nonumber \\
 & & \left. +(\frac{3}{4}\beta_f+12 \gamma_f+\delta_f) m_f^2 (A_Z-B_Z)^2
  \left[ 1-4\left( \frac{m}{m_Z} \right)^2 \right]^2 \right\}
\eea
where
\beq
\beta_f=\frac{m_f^2}{m^2-m_f^2},~~~
\gamma_f=\frac{m^2 (m_Z^2-4 m^2)}{(m_Z^2-4 m^2)^2+\Gamma_Z^2 m_Z^2},~~~
\delta_f=\frac{3}{2}.
\eeq
The value of $a_x$ obtained is identical to the expression for
${\tilde a}_Z$ found by Ellis {\it et al.}  For $b_x$ I find that
\beq
{\tilde b}_Z=b_x -\delta_f a_x=b_x-\frac{3}{2} a_x.
\eeq
This discrepancy can be accounted for by different thermal
averaging prescriptions.  The thermal averaging procedure of
ref.~\cite{srednicki},
which was used by ref.~\cite{ellis},
requires an additional factor of $3a_x/2$ to be subtracted from $b_x$.

\section{Conclusion}

Calculating annihilation cross sections accurately is extremely
important since they, among other things,
sensitively affect cosmological mass
bounds of massive cold relics.  Unfortunately, it is rare to
find two independent calculations in the literature which agree
with each other on a particular annihilation cross section.
The technique outlined in this letter reduces substantially
the possibilities of errors, and also greatly
reduces the amount of effort required to calculate annihilation
cross sections in the low velocity limit.  Furthermore,
using a symbolic manipulation program with this technique reduces
even the most complicated processes to a few seconds of CPU time.

\bigskip
\bigskip
\noindent
{\it Note Added:}  While completing this work I was made aware of
a similar approach to this problem~\cite{leszek}.

\bigskip
\noindent
{\it Acknowledgements:}  I would like to thank G. Kane, C. Kolda,
I. Rothstein, and R. Watkins for reading the manuscript.  I would also
like to thank L.~Roszkowski for useful discussions on thermal averaging.

\vfill\eject

\end{document}